# Counting the Number of Minimum Roman Dominating Functions of a Graph

SHI ZHENG and KOH KHEE MENG, National University of Singapore

We provide two algorithms counting the number of minimum Roman dominating functions of a graph on $n$ vertices in $\mathcal{O}(1.5673^n)$ time and polynomial space. We also show that the time complexity can be reduced to $\mathcal{O}(1.5014^n)$ if exponential space is used. Our result is obtained by transforming the Roman domination problem into other combinatorial problems on graphs for which exact algorithms already exist.

## 1. INTRODUCTION

In algorithmic graph theory, exact exponential-time algorithms are those using exponential executive time to give exact solutions to the problem. This field has grown tremendously recently and many new techniques have been developed, for instance, measure and conquer by Fomin et al. [2005] which is a technique for estimating bounds on the running time. There are also many known results introduced to the field, for instance, inclusion/exclusion [Björklund and Husfeldt. 2006] and dynamic programming on graph decompositions [Fomin and Stepanov. 2007; van Rooij et al. 2009]. Exact exponential-time algorithms on many problems in domination and its variants have been designed and the bounds on running time continue to be improved by newly developed or introduced techniques.

Roman domination is a variant of domination raised by Cockayne et al. [2004] in which the algorithmic and complexity properties of Roman domination was mentioned as an open problem. Many theoretical results in Roman domination have been found and there exist polynomial time algorithms for computing the Roman domination numbers of some special classes of graphs [Cockayne et al. 2004; Liedloff et al. 2005]. However, the Roman domination problem for general graphs is NP-complete as stated in Cockayne et al. [2004], and thus we only expect to construct exact exponential time algorithm on the Roman domination problem. Until now, no exact algorithm has been constructed on the Roman domination problem, and the best result on time complexity is $\mathcal{O}(3^n n^{\mathcal{O}(1)})$ of the trivial algorithm by listing and checking.

*Our Results.* In this paper, we are going to introduce exact exponential-time algorithms counting the number of minimum Roman dominating functions by known results in exact algorithms on domination and its variants. Two algorithms using $\mathcal{O}(1.5673^n)$ time and polynomial space will be introduced. It is noted that the approaches to construct these two algorithms are different. The first algorithm is inspired by using a weight set cover instance to model the Roman domination problem, while the second is constructed by relating Roman domination to partial domination. In addition, an algorithm using $\mathcal{O}(1.5014^n)$ time and exponential space can be derived from the second algorithm.

This paper falls into three sections. In Section 2, we provide some basic definitions and results. In Section 3, the algorithms counting the number of minimum Roman dominating functions are given.

## 2. DEFINITIONS AND BASIC RESULTS

### 2.1 Domination in graphs

Let $G = (V, E)$ be a simple graph with vertex set $V$ and edge set $E$. The *open neighborhood* of a vertex $v$, denoted as $N(v)$, is the set of vertices in $G$ that are adjacent to $v$, and $N[v] = \{v\} \cup N(v)$ is the *closed neighborhood* of $v$.

A subset $D \subseteq V$ is called a *dominating set* of $G$ if for every $v \in V$, $v$ either belongs to $D$ or is adjacent to some vertex in $D$. A dominating set is said to be *minimal* if no proper subset of it is a dominating set of $G$. The minimum cardinality of a dominating set is called the *domination number* of $G$, denoted as $\gamma(G)$, and any dominating set of cardinality $\gamma(G)$ is called a *minimum dominating set* or $\gamma$-*set*. If there is a weight function $w: V \to \mathbb{R}^+$, then $D \subseteq V$ is called a *minimum weight dominating set* if $D$ is a dominating set of the minimum weight. In the case $w: V \to \{1\}$, the minimum weight dominating set problem is reduced to the minimum dominating set problem.

A subset $D \subseteq V$ is called a *partial dominating set* if there are at most $n - l$ vertices that are neither in $D$ nor adjacent to any vertex in $D$. In other words, $D$ dominates at least $l$ vertices.

In a bipartite graph $G = (R \cup B, E)$, $D \subseteq R$ is called a *red-blue dominating set* if $B \subseteq N(D)$ and $D$ is said to be minimum if it has the minimum cardinality.

For a function $f: V \to \{0, 1, 2\}$, $f$ is called a *Roman dominating function*, or abbreviated as RDF, if for every $u \in V_0$, $V_2 \cap N(u)$ is not empty. For a subset $V' \subseteq V$, define $f(V') = \sum_{v \in V'} f(v)$ and the weight of $f$ is $f(V) = \sum_{v \in V} f(v)$. The minimum weight of an RDF of $G$ is the *Roman domination number* of $G$, denoted as $\gamma_R(G)$, and a RDF $f$ is called a *minimum Roman dominating function* or $\gamma_R$-*function* if $w(f) = \gamma_R(G)$. The triple $\{V_0, V_1, V_2\}$ is a vertex partition, where $V_i = \{v \mid f(v) = i\}$ for $i = 0, 1, 2$, and since there is a 1-1 correspondence between functions $f: V \to \{0, 1, 2\}$ and vertex partitions $\{V_0, V_1, V_2\}$, we may write $f = \{V_0, V_1, V_2\}$ for simplicity.

### 2.2 Set cover

A *set cover instance* is defined by a universe $\mathcal{U}$ and a collection $\mathcal{S}$ of subsets such that $\mathcal{U} = \bigcup_{R \in \mathcal{S}} R$. A *set cover* of a set cover instance $(\mathcal{U}, \mathcal{S})$ is a sub-collection $\mathcal{S}' \subseteq \mathcal{S}$ such that $\mathcal{U} = \bigcup_{R \in \mathcal{S}'} R$.

The cardinality of a set cover is the number of subsets in the set cover. A set cover is said to be *minimal* if none of its proper sub-collection is a set cover. A *minimum set cover* is the one of the minimum cardinality. If there is a weight function $w: \mathcal{S} \to \mathbb{R}^+$, then $\mathcal{S}' \subseteq \mathcal{S}$ is called a *minimum weight set cover* if $\mathcal{S}'$ is a set cover of the minimum weight.

A set cover instance $(\mathcal{U}, \mathcal{S})$ can also be formulated as the red-blue dominating set problem on $G = (R \cup B, E)$ by taking $R = \{S \mid S \in \mathcal{S}\}$, $B = \{e \mid e \in \mathcal{U}\}$ and $(e, S) \in E$ if and only if $e \in S$ for any $e \in \mathcal{U}$ and $S \in \mathcal{S}$. $G = (R \cup B, E)$ is called the *incidence graph* of the set cover instance $(\mathcal{U}, \mathcal{S})$. Sometimes the red-blue dominating set problem is considered instead of the set cover problem.

For a graph $G = (V, E)$, if we take $\mathcal{U} = V$ and $\mathcal{S} = \{N[v] \mid v \in V\}$, then $D \subseteq V$ is a minimal or minimum dominating set if and only if $\mathcal{S}' = \{N[v] \mid v \in D\}$ is a minimal or minimum, respectively, set cover of the set cover instance $(\mathcal{U}, \mathcal{S})$. Hence $(\mathcal{U}, \mathcal{S})$ defined above is often used as the set cover modelling of the dominating set problem on $G$. Almost all exponential-time exact algorithms after Gradoni [2006] apply algorithms designed on the set cover problems on the set cover modelling of the dominating set problem.

## 3. ALGORITHMS

In this section, we are going to introduce two exact algorithms on Roman domination. Both of the algorithms count the number of $\gamma_R$-functions of a graph $G$ of $n$ vertices using $\mathcal{O}(1.5673^n)$ time and polynomial space, however they are constructed by different approaches.

The first algorithm is designed by using a weight set cover modelling of the Roman domination problem and the second is designed by relating the Roman domination problem to the partial dominating set problem in graphs. There exist exact algorithms on both weight set cover problems and partial dominating set problems and therefore we will directly include them as subroutines in our algorithms on Roman domination.

In addition, by the second approach, we can also introduce an algorithm using $\mathcal{O}(1.5014^n)$ time and exponential space to count the number of $\gamma_R$-functions of a graph $G$.

### 3.1 An algorithm counting the number of $\gamma_R$-functions

For any simple graph $G = (V, E)$ on $n$ vertices, we introduce a weight set cover instance $(\mathcal{U}_R, \mathcal{S}_R)$, here $\mathcal{U}_R = V$ and $\mathcal{S}_R = \{v \mid v \in V\} \cup \{N[v] \mid v \in V\}$ with the weight function $w(\{v\}) = 1$ and $w(N[v]) = 2$ for each $v \in V$. We notice that for any RDF $f = \{V_0, V_1, V_2\}$ of $G$, $\mathcal{S} = \{v \mid v \in V_1\} \cup \{N[v] \mid v \in V_2\}$ is a set cover of $(\mathcal{U}_R, \mathcal{S}_R)$. And for a set cover $\mathcal{S} = \{v \mid v \in V_1\} \cup \{N[v] \mid v \in V_2\}$ such that $V_1 \cap V_2$ is empty, the function $f = \{V \setminus \{V_1 \cup V_2\}, V_1, V_2\}$ is a RDF. It is not hard to see that for any minimum weight set cover $\mathcal{S} = \{v \mid v \in V_1\} \cup \{N[v] \mid v \in V_2\}$, $V_1 \cap V_2 = \varnothing$; otherwise, if $v \in V_1 \cap V_2$, then $\mathcal{S} \setminus \{v\}$ is a set cover of weight smaller than the weight of $\mathcal{S}$, which is a contradiction. Hence, solving the Roman domination problem on $G$ is equivalent to solving the minimum weight set cover problem on $(\mathcal{U}_R, \mathcal{S}_R)$ as defined above.

Van Rooij [2010] showed that the minimum weight dominating set problem can be solved in $\mathcal{O}(1.5673^n)$ time and polynomial space if the set of possible weight sums is polynomially bounded. In his work, an algorithm listing the number of set covers of each weight $\kappa$ was designed and applied on a weight set cover instance reduced from the minimum weight dominating set problem. Here we can adopt this algorithm to solve our minimum weight set cover problem on $(\mathcal{U}_R, \mathcal{S}_R)$ since the possible weight of a subset is either $1$ or $2$.

---

**ALGORITHM 1.** An Algorithm Computing the Number of $\gamma_R$-functions

**Input:** A simple graph $G = (V, E)$
**Output:** The number of $\gamma_R$-functions(#$\gamma_R$-functions)
**1.** Generate a weight set cover instance $(\mathcal{U}_R, \mathcal{S}_R)$ with weight function $w: \mathcal{U}_R = V$, $\mathcal{S}_R = \{v \mid v \in V\} \cup \{N[v] \mid v \in V\}$ and $w(\{v\}) = 1$ and $w(N[v]) = 2$ for each $v \in V$
**2.** Generate the incidence graph $I(\mathcal{U}_R \cup \mathcal{S}_R, E)$ of $(\mathcal{U}_R, \mathcal{S}_R)$: $(e, S) \in E$ if and only if $e \in S$ for any $e \in \mathcal{U}_R$ and $S \in \mathcal{S}_R$
**3.** Compute the number of set covers of each possible weight: $CWSC(I(\mathcal{U}_R \cup \mathcal{S}_R, E), \varnothing)$
**4.** Return the number of $\gamma_R$-functions: #$\gamma_R$-functions is the number of set covers of the minimum weight

---

**SUBROUTINE** $CWSC$. An Algorithm Counting the Number of Set Covers of Each Weight $\kappa$

**Input:** The incidence graph $I(\mathcal{U} \cup \mathcal{S}, E)$ and a set of annotated vertices $A$
**Output:** A list containing the number of set covers of $(\mathcal{U}, \mathcal{S})$ of each weight $\kappa$

Table I. The weights $v(i)$ and $w(i)$ in complexity analysis of Algorithm 1

| $i$ | 0 | 1 | 2 | 3 | 4 | 5 | 6 | >6 |
|---|---|---|---|---|---|---|---|---|
| $v(i)$ | 0 | 0 | 0.640171 | 0.888601 | 0.969491 | 0.998628 | 1.000000 | 1.000000 |
| $w(i)$ | 0 | 0 | 0.815190 | 1.218997 | 1.362801 | 1.402265 | 1.402265 | 1.402265 |

Source: [Van Rooij 2010]

The subroutine *CWSC* will call itself recursively and the input annotated vertices $A$ is set to be empty initially. For more details of *CWSC* see van Rooij [2010].

THEOREM 3.1. The Roman domination problem can be solved in $\mathcal{O}(1.5673^n)$ time and polynomial space.

PROOF. We consider using Algorithm 1 to solve the Roman domination problem. The main part of Algorithm 1 is obviously in line 3 which is applying *CWSC* on the weight set cover instance $(\mathcal{U}_R, \mathcal{S}_R)$, and the time complexity of Algorithm 1 can be bounded by that of $CWSC(I(\mathcal{U}_R \cup \mathcal{S}_R, E), \varnothing)$. By using measure and conquer technique, van Rooij [2010] showed that the time complexity of *CWSC* can be bounded by $\mathcal{O}(1.205693^{k(\mathcal{S})})$, here

$$k(\mathcal{S}) = \sum_{e \in \mathcal{U}} v(d_{(\mathcal{U} \cup \mathcal{S}) \setminus A}(e)) + \sum_{S \in \mathcal{S}} w(d_{(\mathcal{U} \cup \mathcal{S}) \setminus A}(S))$$

$d_X(u)$ is the degree of $u \in X$ in the induced graph $G[X]$ and $v(i)$ and $w(i)$ are weights set as in Table I.

Hence, for the set cover instance $(\mathcal{U}_R, \mathcal{S}_R)$,

$$k(\mathcal{S}_R) \leq (w(1) + w(>6) + v(>6))n \leq 2.402265n$$

and therefore the time complexity of Algorithm 1 can be bounded from above by $\mathcal{O}(1.205693^{2.402265n}) \leq \mathcal{O}(1.5673^n)$. □

Actually this is also the complexity of applying *CWSC* on the set cover modelling of the minimum dominating set problem since vertices representing subsets of degree 1 in $G[(\mathcal{U} \cup \mathcal{S}) \setminus A]$ do not contribute to the complexity. However, there is a faster version of *CWSC* that uses exponential space in which the vertices representing subsets of degree 1 will contribute to the complexity and the cost of applying it on $(\mathcal{U}_R, \mathcal{S}_R)$ will be exponential space and greater than $\mathcal{O}(1.5673^n)$ time. Hence we are not going to introduce exponential space exact algorithms by using the weight set cover modelling here.

### 3.2 Another algorithm counting the number of $\gamma_R$-functions

Given a simple graph $G = (V, E)$ and a $\gamma_R$-function $f = \{V_0, V_1, V_2\}$ of $G$, for every vertex $u \in V_0$, $u$ is adjacent to at least one vertex $v \in V_2$. Moreover, there is no edge joining $V_1$ and $V_2$. Hence, $V_2$ is a partial dominating set that dominates exactly $n - |V_1|$ vertices and the number of RDF such that $|V_1| = k$ and $|V_2| = j$ and there is no edge joining $V_1$ and $V_2$ is the number of partial dominating sets of cardinality $j$ that dominates exactly $n - k$ vertices.

Recall that for a graph $G = (V, E)$, the set cover modelling of the minimum dominating set problem on $G$ is a set cover instance $(\mathcal{U}, \mathcal{S})$, where $\mathcal{U} = V$ and $\mathcal{S} = \{N[v] \mid v \in V\}$ and the incidence graph $I(\mathcal{U} \cup \mathcal{S}, E)$ of $(\mathcal{U}, \mathcal{S})$ is a bipartite graph such that $(e, S) \in E$ if and only if $e \in S$ for any $e \in \mathcal{U}$ and $S \in \mathcal{S}$. Then $D$ is a partial dominating set of cardinality $j$ that dominates exactly $n-k$ vertices of $G$ if and only if $D$ is a partial red-blue dominating set of $j$ vertices in $\mathcal{S}$ that dominates exactly $n-k$ vertices in $\mathcal{U}$ of the incidence graph $I$. Nederlof and van Rooij [2010] showed that the partial dominating set problem can be solved in $\mathcal{O}(1.5673^n)$ time and polynomial space by an algorithm that computes a matrix $P$ containing the number of partial red-blue dominating sets of $j$ red vertices dominating exactly $l$ blue vertices for each given $0 \leq j \leq |R|$ and $0 \leq l \leq |B|$ for any bipartite graph $G = (R \cup B, E)$. Here we will include this algorithm as a subroutine.

---

**ALGORITHM 2.** An Algorithm Computing the Number of $\gamma_R$-functions

---

**Input:** A simple graph $G = (V, E)$
**Output:** The number of $\gamma_R$-functions(# $\gamma_R$-functions)
**1.** Generate a set cover instance $(\mathcal{U}, \mathcal{S})$: $\mathcal{U} = V$ and $\mathcal{S} = \{N[v] \mid v \in V\}$
**2.** Generate the incidence graph $I(\mathcal{U} \cup \mathcal{S}, E)$ of $(\mathcal{U}, \mathcal{S})$: $(e, S) \in E$ if and only if $e \in S$ for any $e \in \mathcal{U}$ and $S \in \mathcal{S}$
**3.** Compute a matrix $P$ containing the number of partial red-blue dominating sets of cardinality $j$ that dominates exactly $l$ vertices: $CPSC(I(\mathcal{U} \cup \mathcal{S}, E), \varnothing)$
**4.** Compute a matrix $A$ containing the number of RDF such that $|V_1| = k$ and $|V_2| = j$ and there is no edge joining $V_1$ and $V_2$: $a_{jk} = p_{j, n-k}$
**5.** Compute the Roman domination number: $\gamma_R = \min_{a_{jk} > 0} \{2j + k\}$
**6.** Compute the number of $\gamma_R$-functions: $\# \gamma_R\text{-functions} = \sum_{2j+k=\gamma_R} a_{jk}$

---

**SUBROUTINE** *CPSC*. An Algorithm Computing the Number of Partial Red-Blue Dominating Sets

---

**Input:** A bipartite graph $G = (R \cup B, E)$ and a set of annotated vertices $A$
**Output:** A matrix containing the number of partial red-blue dominating sets of cardinality $j$ that dominates exactly $l$ vertices for each given $0 \leq j \leq |R|$ and $0 \leq l \leq |B|$

---

The subroutine *CPSC* is very similar to *CWSC* as the inputs are both a bipartite graph and a set of annotated vertices that is empty initially and the outputs can be generated by similar rules. This similarity leads to that the complexity analysis of *CPSC* is the same as that of *CWSC*.

AN ALTERNATE PROOF FOR THEOREM 3.1. We consider using Algorithm 2 to solve the Roman domination problem. The main part of Algorithm 2 is line 3 which applies *CPSC* on the weight set cover instance $(\mathcal{U}, \mathcal{S})$ and the time complexity of Algorithm 2 can be bounded by that of $CPSC(I(\mathcal{U} \cup \mathcal{S}, E), \varnothing)$. The complexity was analyzed to be $\mathcal{O}(1.5673^n)$ time and polynomial space [Nederlof and van Rooij. 2010]. □

There is an exponential-space version of *CPSC* using $\mathcal{O}(1.5014^n)$ time solving the partial dominating set problem. For more details see Nederlof and van Rooij [2010].

THEOREM 3.2. The Roman domination problem can be solved in $\mathcal{O}(1.5014^n)$ time and exponential space.

PROOF. We consider substituting *CPSC* in Algorithm 2 with its exponential-space version and use it to solve the Roman domination problem. The time complexity can be estimated in the way similar to the alternate proof for Theorem 3.1 to be $\mathcal{O}(1.5014^n)$. □

**REFERENCES**

Björklund, A. and Husfeldt, T. 2006. Inclusion--Exclusion Algorithms for Counting Set Partitions. In *Foundations of Computer Science, 2006. FOCS '06. 47th Annual IEEE Symposium on*, 575-582.

Cockayne, E.J., Dreyer, P.A., Hedetniemi, S.M. and Hedetniemi, S.T. 2004. Roman domination in graphs. *Discrete Mathematics 278*, 11-22.

Fomin, F., Grandoni, F. and Kratsch, D. 2005. Measure and Conquer: Domination – A Case Study. In *Automata, Languages and Programming*, L. CAIRES, G. ITALIANO, L. MONTEIRO, C. PALAMIDESSI AND M. YUNG Eds. Springer Berlin Heidelberg, 191-203.

Fomin, F. and Stepanov, A. 2007. Counting Minimum Weighted Dominating Sets. In *Computing and Combinatorics*, G. LIN Ed. Springer Berlin Heidelberg, 165-175.

Grandoni, F. 2006. A note on the complexity of minimum dominating set. *Journal of Discrete Algorithms 4*, 209-214.

Liedloff, M., Kloks, T., Liu, J. and Peng, S.-L. 2005. Roman Domination over Some Graph Classes. In *Graph-Theoretic Concepts in Computer Science*, D. KRATSCH Ed. Springer Berlin Heidelberg, 103-114.

Nederlof, J. and van Rooij, J.M.M.V. 2010. *Inclusion/Exclusion Branching for Partial Dominating Set and Set Splitting*.

Van Rooij, J.M. 2010. Polynomial Space Algorithms for Counting Dominating Sets and the Domatic Number. In *Algorithms and Complexity*, T. CALAMONERI AND J. DIAZ Eds. Springer Berlin Heidelberg, 73-84.

Van Rooij, J.M., Bodlaender, H. and Rossmanith, P. 2009. Dynamic Programming on Tree Decompositions Using Generalised Fast Subset Convolution. In *Algorithms - ESA 2009*, A. FIAT AND P. SANDERS Eds. Springer Berlin Heidelberg, 566-577.